\documentclass[manuscript]{aastex}
\usepackage{emulateapj5}

\newcommand{\lsim}{\raisebox{-0.3ex}{\mbox{$\stackrel{<}{_\sim} \,$}}}

\usepackage{epsfig}
\usepackage{amsmath}
\usepackage{amssymb}
\usepackage{natbib}
\usepackage{graphicx}

\shorttitle{X-ray bursts from IGR J17480--2446}
\shortauthors{Chakraborty and Bhattacharyya}
\begin{document}

\title{X-ray bursts from the Terzan 5 transient IGR J17480--2446: nuclear rather than gravitational} 

\author{Manoneeta Chakraborty\altaffilmark{1} and Sudip Bhattacharyya\altaffilmark{1}}
\altaffiltext{1}{Tata Institute of Fundamental Research,
    Mumbai-400005, India; manoneeta@tifr.res.in; sudip@tifr.res.in}

\begin{abstract}
The 2010 outburst of the transient neutron star low-mass X-ray binary
IGR J17480--2446 has exhibited a series of unique X-ray bursts, as well
as millihertz (mHz) quasi-periodic oscillations (QPOs) related to these
bursts. It has been recently proposed that these are type-II bursts,
powered by the gravitational energy. This implies that the current
nuclear-burning based model of mHz QPOs is not correct, and this
timing feature cannot be used as a tool to measure the neutron star 
parameters. We report the analysis of the {\it Rossi X-ray Timing Explorer}
data of IGR J17480--2446 to show that the burst properties of this source
are quite different from the properties of the type-II bursts observed
from the rapid burster and GRO J1744--28. For example, the 
inferred ratio ($\sim 50-90$) of the non-burst fluence to burst fluence
is consistent with the thermonuclear origin of IGR J17480--2446 bursts,
and is significantly different from this ratio ($\lsim 4$) for type-II bursts.
Our results suggest that the bursts and the mHz QPOs from 
IGR J17480--2446 are powered by the nuclear energy.
\end{abstract}

\keywords{accretion, accretion disks --- stars: neutron --- X-rays: binaries --- 
X-rays: bursts  --- X-rays: individual: IGR J17480-2446 --- X-rays: stars}

\section{Introduction}\label{Introduction}

The first observed outburst \citep{Bordasetal2010} 
of the globular cluster Terzan 5 source IGR J17480--2446 in Oct/Nov, 2010
revealed that it is a neutron star low-mass X-ray binary (LMXB) and
an 11 Hz pulsar \citep{Chenevezetal2010, StrohmayerMarkwardt2010}.
An X-ray burst was observed from this source with {\it Rossi X-ray Timing Explorer}
({\it RXTE}) on Oct 13, 2010 \citep{StrohmayerMarkwardt2010}. While this is
not surprising even for a pulsar, what made this source very interesting is the
recurrent bursts observed for next few days \citep{Altamiranoetal2010b, Papittoetal2011}.
As the source intensity increased, the bursts became more frequent, and eventually
disappeared, and millihertz (mHz) quasi-periodic oscillations (QPOs) appeared
\citep{Altamiranoetal2010b, Linaresetal2010}. Finally, the bursts reappeared,
and as the source intensity decreased, they became less frequent 
\citep{ChakrabortyBhattacharyya2010}. The burst recurrence frequencies
observed between October 13 and 16 converge asymptotically towards the mHz QPO 
frequency, which suggests that these QPO is related to bursts \citep{Linaresetal2010}.
If these bursts are type-I or thermonuclear \citep{StrohmayerBildsten2006}, then such
a relation is expected from a widely accepted mHz QPO model, which 
interprets this timing feature as a signature of marginally stable nuclear burning
\citep{Hegeretal2007}. However, \citet{GallowayZand2010} reported that, except
the Oct 13 burst, no other burst up to Oct 26 showed a clear cooling trend
during burst decay.
This motivated them to suggest that, except the first burst, others are 
type-II bursts.
The type-II bursts are believed to be caused by the accretion instability, and hence 
powered by the gravitational potential energy. 
This would mean that the mHz QPOs are related to the accretion dynamics, rather than
nuclear burning, and hence plausibly could not be used as a tool to measure the neutron star
parameters \citep{Hegeretal2007, Bhattacharyya2010}.
Type-II bursts have so far been observed from
two sources: rapid burster and GRO J1744--28 \citep{Tanetal1991, Lewinetal1995, Lewinetal1996}.
\citet{GallowayZand2010} have also indicated that IGR J17480--2446 might be
a GRO J1744--28 analogue (since both have relatively wide orbit and slow pulsations),
further strengthening the gravitational origin argument for the IGR J17480--2446 bursts.
It is therefore extremely important to investigate whether
the X-ray bursts from IGR J17480--2446 are powered by the nuclear energy or
by the gravitational energy (1) in order to understand the  nature of 
the unique source IGR J17480--2446, and its unique bursts, and (2) to find 
out the origin of mHz QPOs and their usefulness in understanding thermonuclear bursts and 
in measuring neutron star parameters.

We have performed spectral and timing analysis of the entire RXTE 
Proportional Counter Array (PCA) data 
from IGR J17480--2446, the details of which will be reported elsewhere. 
In this Letter, we report and discuss the results required to address the
single very important question mentioned above.

\section{Observation and Data Analysis}\label{ObservationandDataAnalysis}

The neutron star LMXB IGR J17480--2446 was observed almost everyday with 
{\it RXTE} between Oct 13, 2010 and Nov 19, 2010, i.e., 
during its outburst. The total observation time was $\approx 297$ ks (proposal no. 95437;
46 obsIds: 95437-01-01-00 to 95437-01-14-00). 
We have found $\sim 400$ bursts (including the very weak ones) 
in the {\it RXTE} data of this source, and 
analyzed the entire PCA data with the aim of understanding
the nature of these unique bursts. In order to examine the lightcurves, and to measure
the source properties such as burst duration, burst interval (time gap between two successive
bursts) and spectral evolution during a burst, we have used the event-mode/good-Xenon files.
This is because, this mode has sufficiently good time resolution
and spectral resolution. In order to perform the time-resolved spectroscopy
of the bursts, which could be comfortably distinguished from the non-burst
emissions, we have divided each burst into segments with sufficient statistics.
For each segment, we have created a total energy spectrum
with deadtime correction \citep{vanderKlis1989}, and a background spectrum from the pre-burst
emission \citep{BhattacharyyaStrohmayer2006,
Gallowayetal2008}. Usually a blackbody model is used to describe a burst spectrum, and hence 
we have fitted each spectrum in $3-15$ keV with an absorbed blackbody
({\tt phabs*bbodyrad} in XSPEC) for a fixed neutral hydrogen column density
$N_{\rm H} = 3.8\times10^{22}$ cm$^{-2}$ \citep{GallowayZand2010, Kuulkersetal2003}.
The blackbody provides acceptable fits (for the degrees of freedom $\nu = 17$) 
with $\chi^2/\nu \le 1.0$ for $\sim 40$\% segments and 
$1 < \chi^2/\nu \le 1.5$ for $\sim 45$\% segments, 
excluding a few segments with $\chi^2/\nu > 2.0$.
We have used these fits to infer the total energy (time-integrated flux or
fluence), as well as the evolution of
temperature, emitting area and flux of each burst. We have also fitted the deadtime
corrected non-burst spectra in $3-15$ keV, using backgrounds created from
the {\it RXTE} data analysis tool {\tt PCABACKEST}. These spectra were made 
from standard-2 mode files.
First we have tried the conventionally used thermal models absorbed blackbody
({\tt phabs*bbodyrad} in XSPEC) and absorbed disk-blackbody ({\tt phabs*diskbb}
in XSPEC), as well as the non-thermal model absorbed powerlaw ({\tt phabs*powerlaw} 
in XSPEC), in order to fit the non-burst spectra. Similar to the burst spectral
fitting, we have fixed $N_{\rm H}$ at $3.8\times10^{22}$ cm$^{-2}$. Each of these
single component models gave extremely bad fit. Next we have tried three two-component
models combining two of {\tt bbodyrad}, {\tt diskbb} and {\tt powerlaw} in rotation.
The {\tt phabs*(bbodyrad+powerlaw)} model gave much better fit than {\tt phabs*(bbodyrad+diskbb)},
and somewhat better fit than {\tt phabs*(diskbb+powerlaw)}. Furthermore, the addition of a Gaussian 
component describing an iron K$\alpha$ emission 
line between $6.4-6.97$ keV improves the fit significantly (e.g., significance
$\approx 1-5.6\times10^{-4}$ for the Oct 26, 2010 data), and gives acceptable fits
(see also \citet{Chakrabortyetal2011}). 
Therefore, we have fitted each non-burst spectrum with the 
{\tt phabs*(bbodyrad+powerlaw+Gaussian)} model in order to infer the flux, fluence 
and other spectral parameters. 
We have tried to estimate the contributions of the field sources to this non-burst
flux. The total 2--6 keV {\it Chandra} count rate of all the sources 
(except CX25, which is IGR J17480--2446; \citet{Pooleyetal2010}) given in Tables 1 and 2 
of \citet{Heinkeietal2006} is $\approx 0.06$ cps, which is about 0.1\% of the 2--6 keV 
{\it Chandra} count rate equivalent to the lowest (of Oct 13, 2010) observed {\it RXTE}
PCA flux. This shows that the non-burst flux has been overestimated by $\lsim 0.1\%$.
 
\section{Results and Discussion}\label{ResultsandDiscussion}

In this Letter, we report the evolution of various burst and non-burst properties
of IGR J17480--2446 throughout the source outburst. 
Examples of a few days given in Table~\ref{Properties} and Fig.~\ref{lc}
show that the burst properties are well correlated with the non-burst intensity.
The Oct 13 burst detected with {\it RXTE} was likely to be powered by nuclear energy 
because of the following reasons.
(a) The burst light curve rises rapidly and decays slowly like a typical
thermonuclear burst (Fig.~\ref{lc}). (b) The blackbody temperature decreases during
burst decay \citep{GallowayZand2010}. (c) Intensity variation
of this burst at the pulsar frequency was detected \citep{Altamiranoetal2010a},
and this feature is broader in power spectrum during the burst than in the non-burst
portions. Here we discuss the properties of the bursts following the Oct 13 burst,
in order to find out whether these bursts are powered by nuclear energy or 
gravitational energy.

(1) Burst spectrum vs. non-burst spectrum:
The burst spectrum of IGR J17480--2446 is typically well described with
a blackbody (\S~\ref{ObservationandDataAnalysis}), and the addition of a 
powerlaw does not improve the fit (note that, for fainter bursts, this might
be due to insufficient statistical quality). The non-burst spectrum of IGR J17480--2446 
cannot be fitted with a blackbody, and a blackbody plus powerlaw model
is required (see \S~\ref{ObservationandDataAnalysis}; see also \citet{Chakrabortyetal2011}). So these two
spectra are plausibly sufficiently different, indicating different origins.
However, these two spectra are similar to each other for GRO J1744--28
\citep{Lewinetal1996}.

(2) Temperature evolution during burst decay:
Most of the bursts from IGR J17480--2446 do not exhibit a clear cooling trend
during the burst decay. However, this cannot rule out the nuclear
origin of the bursts for the following reasons. 
(a) The Nov 18 burst 
shows a cooling trend with a significance of $\approx 1-9\times10^{-8}$,
while the similar bursts of Nov 19 do not show such a trend.
(b) Several likely thermonuclear bursts (e.g., 18th burst of EXO 0748--676 in
\citet{Gallowayetal2008}) did not show a cooling trend.
(c) If the color factor $f$ increases with the decrease of the 
effective temperature $T_{\rm eff}$ (may be possible
for high metalicity of burning matter, see Table 2 of \citet{Majczynaetal2005}),
the cooling trend may not be visible, as the color temperature 
$T_{\rm c} = f T_{\rm eff}$.

(3) Ratio of non-burst fluence to burst fluence:
For a thermonuclear burst, the ratio of the nuclear energy generated
to the gravitational energy released is expected to be $\sim 40$
\citep{StrohmayerBildsten2006}. The ratio of non-burst fluence to burst fluence
gives an upper limit of this energy ratio, as stable nuclear burning
can happen during the non-burst time. The ratio of the fluences ($\sim 50-90$) inferred
for IGR J17480--2446 bursts throughout the outburst is very consistent
with the plausible thermonuclear origin of these bursts (see Table~\ref{Properties}).
However, this ratio is significantly smaller than 40 for rapid burster
($\lsim 1$; \citet{Tanetal1991}) and GRO J1744--28 ($\lsim 4$; \citet{Lewinetal1996}),
indicating that the type-II bursts from these sources are not of 
thermonuclear origin. Note that, although we have estimated the ratio of the fluences 
for IGR J17480--2446 in the $3-15$ keV range, the bolometric ratio
would also be consistent with the expected value $\sim 40$
because of the following reasons. 
(a) Since the non-burst spectrum has two continuum components
(a blackbody similar to the burst spectrum, plus a powerlaw),
a larger portion of the non-burst energy is likely to be outside
the $3-15$ keV range compared to the burst energy.
(b) Even if the entire non-burst energy is in the $3-15$ keV range
(which is the conservative scenario), our estimated ratio will be
less by a small factor of $\approx 1.2$ for a typical burst blackbody
temperature of 2 keV.

(4) Burst shape and duration:
the bursts neither show ``ringing" nor ``flat top" like the type-II
bursts of the rapid burster \citep{Tanetal1991}. Their duration
gradually changes (Table~\ref{Properties}) unlike the GRO J1744--28
bursts \citep{Fishmanetal1996}. Fig.~\ref{shape} clearly shows that the
shapes of the bursts are similar to the that of the Oct 13
burst, and any change is gradual throughout the outburst. 
This would not happen if the Oct 13 burst and the other bursts 
originated from two independent physical mechanisms.
These points support the nuclear origin of the IGR J17480--2446 bursts.

(5) Intensity dips after bursts:
while type-II bursts from the rapid burster and GRO J1744--28 are
followed by intensity dips \citep{Tanetal1991, Lewinetal1996}, no clear evidence
of such dips has been found for IGR J17480--2446 bursts. Only 10\% 
bursts are followed by weak dips, which are plausibly of the non-burst origin.

(6) Burst fluence versus burst interval: burst fluence is positively 
correlated with burst interval for IGR J17480--2446 (Table~\ref{Properties}), while 
the fluence of GRO J1744--28 type-II bursts remained approximately
constant when the burst interval changed by factors of $\sim 4$
\citep{Kouveliotouetal1996}.

(7) Burst peak flux and burst fluence versus non-burst flux:
while with the increase of non-burst flux, both burst peak flux and burst fluence 
of GRO J1744--28 increase \citep{Lewinetal1996}, these parameter values of 
IGR J17480--2446 decrease (Table~\ref{Properties}).

\section{Conclusion}\label{Conclusion}

\begin{figure}[h]
\centering
\includegraphics*[width=7.5cm]{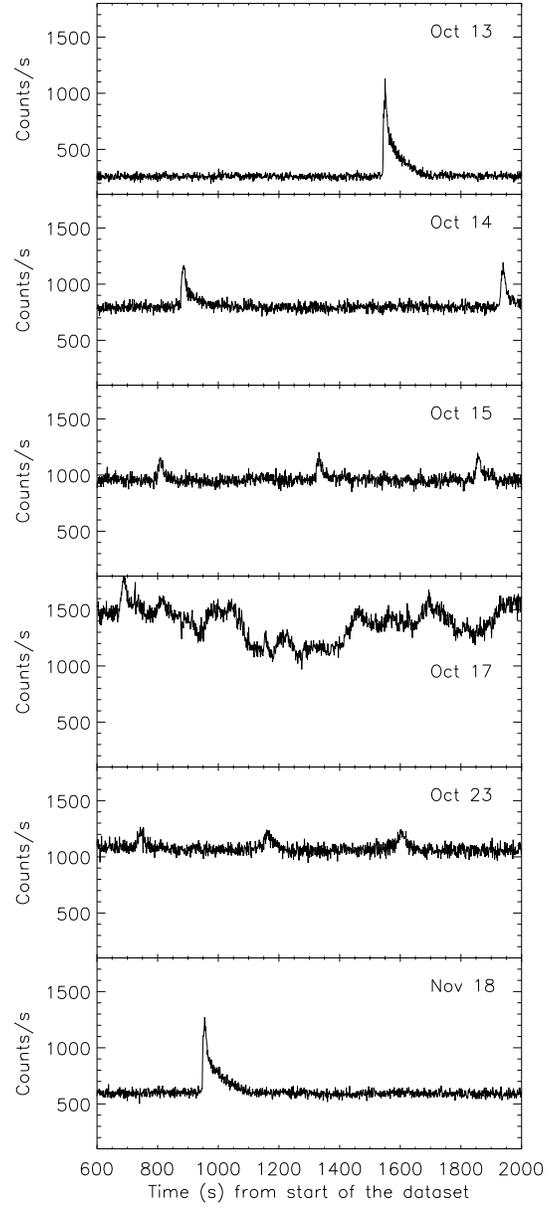}
\caption{Light curves of 1 s resolution of portions of six {\it RXTE} PCU-2 
continuous data sets from IGR J17480--2446. The date of observation of each 
data set is mentioned. This figure shows that during the rise of the 2010 outburst,
as the non-burst flux increases, the burst peak flux and the burst interval
gradually decrease. The initial burst properties return as the source intensity
decays (\S~\ref{ResultsandDiscussion}).
\label{lc}}
\end{figure}

\begin{figure}[h]
\centering
\includegraphics*[width=7.5cm]{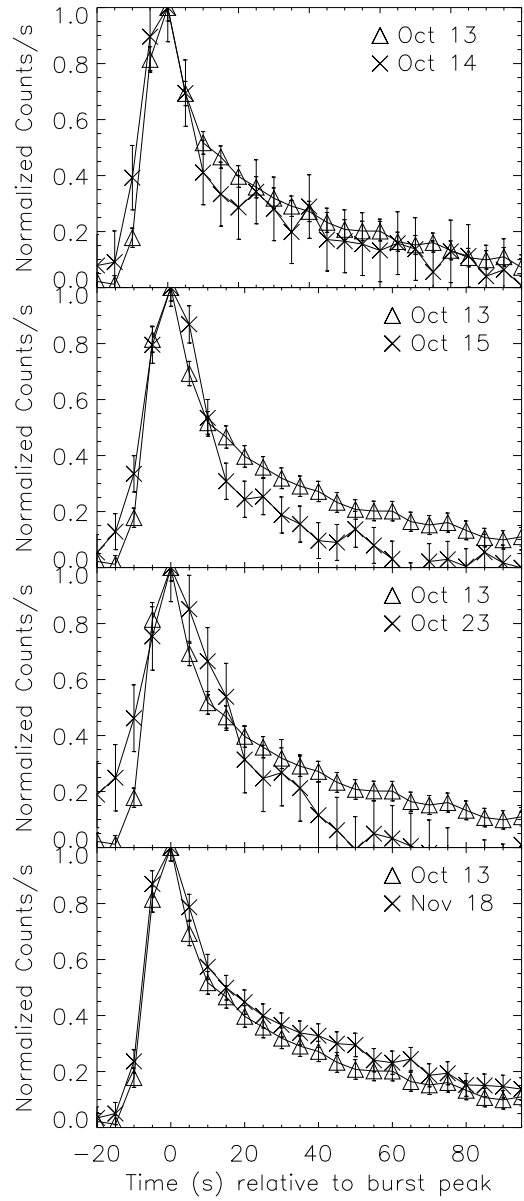}
\caption{Comparisons of the bursts of Oct 14, Oct 15, Oct 23 and Nov 18 with the Oct 13 burst.
Each burst light curve is pre-burst level subtracted and normalized with the peak count rate.
In case of Oct 15, four burst light curves are combined to make the statistics better.
In case of Oct 23, six burst light curves are combined. This figure shows that the shapes
of the bursts of different days are not drastically different from each other, and the gradual change
of the shape is correlated with various burst and non-burst properties (compare with 
Table~\ref{Properties} and Fig.~\ref{lc}). 
\label{shape}}
\end{figure}

The gradual (as opposed to abrupt) change of all the burst properties throughout the outburst
(Table~\ref{Properties} and Fig.~\ref{lc}) strongly indicate that
all the bursts from IGR J17480--2446, including the Oct 13 burst, 
originate from the same physical mechanism.
The points 3, 4 and 5 of \S~\ref{ResultsandDiscussion} suggest that 
these bursts are of thermonuclear origin, and
mHz QPOs, plausibly caused by the quasi-stable nuclear burning, can be
used as a tool to measure the neutron star parameters.
The point 2 of \S~\ref{ResultsandDiscussion} indicates that a clear cooling
may not always be present during a thermonuclear burst decay. 
The points 1, 3, 4, 5, 6 and 7 of \S~\ref{ResultsandDiscussion} 
show that IGR J17480--2446 is not a GRO J1744--28
analogue (\S~\ref{Introduction}).
Finally, we note that the IGR J17480--2446 bursts are similar to the GS 1826-238
bursts in their clock-like recurrence times and in the shortening of the recurrence 
time with the non-burst flux increase \citep{StrohmayerBildsten2006}.\footnotemark[1]

\acknowledgments

We thank Arunava Mukherjee, Jon Miller and Tod Strohmayer for discussion, and an
anonymous referee for constructive comments. This work was supported in part by 
the TIFR plan project 11P-408 (P.I.:K. P. Singh).

\footnotesize{$^1$After this Letter was submitted and made public, the following papers on the bursts
from IGR J17480--2446 were submitted in the arXiv: \citet{Cavecchietal2011}, \citet{Linaresetal2011}. These
papers supported our conclusion regarding the thermonuclear origin of these bursts.}

{}

\clearpage
\begin{table*}
 \centering
\caption{Properties (with 90\% errors) of burst and non-burst emissions during the 2010 outburst
of IGR J17480--2446 (see \S~\ref{ResultsandDiscussion}).
}
\begin{tabular}{ccccccccc}

\hline
Date\footnotemark[1] & Time & Non-burst & Burst & $\tau$ \footnotemark[5] & T$_{int}$\footnotemark[6] & Burst & Ratio of& No. of\\
 & range\footnotemark[2] & flux\footnotemark[3] & peak flux\footnotemark[4] &  &  & fluence\footnotemark[7] & fluences\footnotemark[8] & bursts\footnotemark[9]\\
\hline
Oct 13 & 00:12:32-01:05:10 &  2.58$_{-0.02}^{+0.01}$ & 9.89$_{-0.78}^{+0.73}$ & 120 &    - & 26.70$_{-0.54}^{+0.52}$ &   - &  1\\
Oct 14 & 04:28:01-05:24:09 &  8.62$_{-0.04}^{+0.04}$ & 5.36$_{-0.56}^{+0.48}$ & 105 & 1034 & 15.35$_{-0.67}^{+0.61}$ &  58.11$_{-2.55}^{+2.32}$ &  3\\
Oct 15 & 10:24:32-11:06:11 & 10.65$_{-0.04}^{+0.04}$ & 2.73$_{-0.23}^{+0.02}$ &  72 &  512 &  6.60$_{-0.29}^{+0.22}$ &  82.62$_{-3.59}^{+2.76}$ &  5\\
Oct 23 & 03:26:40-04:14:09 & 11.81$_{-0.06}^{+0.05}$ & 2.70$_{-0.39}^{+0.29}$ &  56 &  434 &  6.72$_{-0.44}^{+0.37}$ &  76.38$_{-4.99}^{+4.25}$ &  6\\
Oct 26 & 08:19:28-09:15:11 & 10.55$_{-0.05}^{+0.04}$ & 4.21$_{-0.31}^{+0.27}$ &  60 &  722 & 11.21$_{-0.35}^{+0.31}$ &  67.94$_{-2.15}^{+1.92}$ &  5\\
Oct 31 & 09:05:20-10:02:10 &  9.54$_{-0.04}^{+0.04}$ & 4.18$_{-0.42}^{+0.38}$ &  90 & 1016 & 10.31$_{-0.42}^{+0.37}$ &  93.96$_{-3.81}^{+3.43}$ &  3\\
Nov 05 & 06:43:28-07:40:10 &  8.92$_{-0.04}^{+0.04}$ & 4.56$_{-0.46}^{+0.41}$ & 100 & 1273 & 12.50$_{-0.38}^{+0.35}$ &  90.83$_{-2.82}^{+2.58}$ &  3\\
Nov 08 & 05:17:20-06:14:10 &  8.35$_{-0.04}^{+0.04}$ & 4.92$_{-0.60}^{+0.54}$ & 120 & 1488 & 17.62$_{-0.46}^{+0.42}$ &  70.48$_{-1.85}^{+1.72}$ &  2\\
Nov 15 & 03:32:32-04:29:09 &  7.26$_{-0.03}^{+0.03}$ & 7.65$_{-0.63}^{+0.59}$ & 130 & 2137 & 26.85$_{-0.59}^{+0.55}$ &  57.80$_{-1.30}^{+1.20}$ &  2\\
Nov 18 & 02:06:24-03:03:11 &  6.62$_{-0.02}^{+0.02}$ & 9.17$_{-0.65}^{+0.61}$ & 135 & 2391 & 36.41$_{-0.70}^{+0.65}$ &  43.49$_{-0.85}^{+0.79}$ &  2\\
\hline
\end{tabular}
\begin{flushleft}
$^1$Date of observation.\\
$^2$Start and end times (UTC) of observation.\\
$^3$Average flux (in $10^{-9}$ ergs cm$^{-2}$ s$^{-1}$) within $3-15$ keV in the portions of the data without bursts.\\
$^4$Average flux (in $10^{-9}$ ergs cm$^{-2}$ s$^{-1}$) within $3-15$ keV of the peak of a burst.\\
$^5$Average duration (in s) of a burst (from the time corresponding to $\sim5$\% of peak count 
rate during rise to the time corresponding to $\sim5$\% of peak count rate during decay).\\
$^6$Average time gap (burst interval; in s) between two successive bursts.\\
$^7$Average total energy (in $10^{-8}$ ergs cm$^{-2}$) within $3-15$ keV of a burst.\\
$^8$Ratio of the non-burst fluence to the corresponding burst fluence.\\
$^9$Number of bursts within the given time range.\\
\end{flushleft}
\label{Properties}
\end{table*}

\clearpage

\clearpage


\begin{thebibliography}{}

\bibitem[Altamirano et al. (2010b)]{Altamiranoetal2010b} Altamirano, D., Homan, J., 
Linares, M., et al. 2010b, ATel, 2952, 1

\bibitem[Altamirano et al. (2010a)]{Altamiranoetal2010a} Altamirano, D., Watts, A.,
 Kalamkar, M., et al. 2010a, ATel, 2932, 1

\bibitem[Bhattacharyya (2010)]{Bhattacharyya2010} Bhattacharyya, S. 2010,
Advances in Space Research, 45, 949

\bibitem[Bhattacharyya and Strohmayer (2006)]{BhattacharyyaStrohmayer2006} Bhattacharyya, S., \& 
Strohmayer, T. E. 2006, ApJ, 641, L53

\bibitem[Bordas et al. (2010)]{Bordasetal2010} Bordas, P., Kuulkers, E., Alfonso-Garz\'on, J.,
et al. 2010, ATel, 2919, 1

\bibitem[Cavecchi et al. (2011)]{Cavecchietal2011} Cavecchi, Y., Patruno, A., Haskell, B., Watts, A. L.,
 Levin, Y., Linares, M., Altamirano, D., Wijnands, R., \& van der Klis, M., 2011, arXiv:1102.1548

\bibitem[Chakraborty and Bhattacharyya (2010)]{ChakrabortyBhattacharyya2010}
Chakraborty, M., \& Bhattacharyya, S. 2010, ATel, 3044, 1

\bibitem[Chakraborty et al. (2011)]{Chakrabortyetal2011} Chakraborty, M., Bhattacharyya, S.,
\& Mukherjee, A. 2011, arXiv:1102.1033v1

\bibitem[Chenevez et al. (2010)]{Chenevezetal2010} Chenevez, J., Kuulkers, E., Alfonso-Garz\'on, J.,
et al. 2010, ATel, 2924, 1

\bibitem[Fishman et al. (1996)]{Fishmanetal1996} Fishman, G. J., et al. 1996, IAU Circ., No. 6290

\bibitem[Galloway and in't Zand (2010)]{GallowayZand2010} Galloway, D. K., \& in't Zand, J. J. M.
2010, ATel, 3000, 1

\bibitem[Galloway et al. (2008)]{Gallowayetal2008} Galloway, D. K., Muno, M. P., Hartman, J. M., 
Psaltis, D., \& Chakrabarty, D. 2008, ApJS, 179, 360

\bibitem[Heger et al. (2007)]{Hegeretal2007} Heger, A., Cumming, A., \&
Woosley, S. E. 2007, ApJ, 665, 1311

\bibitem[Heinke et al. (2006)]{Heinkeietal2006} Heinke, C. O., Wijnands, R., 
Cohn, H. N., Lugger, P. M., Grindlay, J. E., Pooley, D., \& Lewin, W. H. G. 2006, ApJ, 651, 1098

\bibitem[Kouveliotou et al. (1996)]{Kouveliotouetal1996} Kouveliotou, C., van Paradijs, J.,
Fishman, G. J., et al. 1996, Nature, 379, 399

\bibitem[Kuulkers et al. (2003)]{Kuulkersetal2003} Kuulkers, E., den Hartog, P. R., 
in't Zand, J. J. M., Verbunt, F. W. M., Harris, W. E., \& Cocchi, M. 2003, A\&A, 399, 663

\bibitem[Lewin et al. (1996)]{Lewinetal1996} Lewin, W. H. G., Rutledge, R. E.,
Kommers, J. M., van Paradijs, J., \& Kouveliotou, C. 1996, ApJ, 462, L39

\bibitem[Lewin et al. (1995)]{Lewinetal1995} Lewin, W. H. G., van Paradijs, J., \& Taam, R. E.
1995, in X-Ray Binaries, ed. W. H. G. Lewin, J. van Paradijs, \& E. P. J. van den Heuvel 
(Cambridge: Cambridge Univ. Press), 175

\bibitem[Linares et al. (2010)]{Linaresetal2010} Linares, M., Altamirano, D., Watts, A., 
et al. 2010, ATel, 2958, 1

\bibitem[Linares et al. (2011)]{Linaresetal2011} Linares, M., Chakrabarty, D., van der Klis, M., arXiv:1102.1455

\bibitem[Majczyna et al. (2005)]{Majczynaetal2005} Majczyna, A., Madej, J.,
Joss, P. C., \& R\'oaska, A., 2005, A\&A, 430, 643

\bibitem[Papitto et al. (2011)]{Papittoetal2011} Papitto, A., D'A\'i, A., Motta, S., et al. 2011,
A\&A, 526, L3

\bibitem[Pooley et al. (2010)]{Pooleyetal2010} Pooley, D., Homan, J., Heinke, C., Linares, M.,
Altamirano, D., \& Lewin W. 2010, ATel, 2974, 1

\bibitem[Strohmayer and Bildsten (2006)]{StrohmayerBildsten2006}
Strohmayer, T.E., \& Bildsten, L. 2006, in Compact
Stellar X-ray Sources, ed. W.H.G. Lewin and M. van der Klis,
(Cambridge University Press: Cambridge), 39, 113

\bibitem[Strohmayer and Markwardt (2010)]{StrohmayerMarkwardt2010} Strohmayer, T. E., \& Markwardt, C. B.
2010, ATel, 2929, 1

\bibitem[Tan et al. (1991)]{Tanetal1991} Tan, J., Lewin, W. H. G., Lubin, L. M., et al.
1991, MNRAS., 251, 1

\bibitem[{van der Klis}(1989)]{vanderKlis1989} van der Klis, M. 1989, in Timing Neutron Stars, ed. H. \"Ogelman and E. P. J. van den Heuvel, (Kluwer Academic / Plenum Publishers: New York), 27

\end{thebibliography}
\end{document}